\documentstyle[prd,aps,preprint,tighten,epsfig]{revtex}

\begin{document}

\draft

\title{Impacts of the Higgs mass on vacuum stability,
running fermion masses and two-body Higgs decays}
\author{{\bf Zhi-zhong Xing}~$^a$
\thanks{E-mail: xingzz@ihep.ac.cn},
~ {\bf He Zhang}~$^b$
\thanks{E-mail: he.zhang@mpi-hd.mpg.de},
~ {\bf Shun Zhou}~$^c$
\thanks{E-mail: zhoush@mppmu.mpg.de}}
\address{
$^a${\sl Institute of High Energy Physics, Chinese Academy of
Sciences, Beijing 100049, China} \\
$^b${\sl Max-Planck-Institut f\"{u}r Kernphysik,
69029 Heidelberg, Germany} \\
$^c${\sl Max-Planck-Institut f\"{u}r Physik
(Werner-Heisenberg-Institut), 80805 M\"{u}nchen, Germany}}

\maketitle

\begin{abstract}
The recent results of the ATLAS and CMS experiments indicate
$116~{\rm GeV} \lesssim M^{}_H \lesssim 131~{\rm GeV}$ and $115
~{\rm GeV} \lesssim M^{}_H \lesssim 127~{\rm GeV}$, respectively,
for the mass of the Higgs boson in the standard model (SM) at the
$95\%$ confidence level. In particular, both experiments point to a
preferred narrow mass range $M^{}_H \simeq (124 \cdots 126)~{\rm
GeV}$. We examine the impact of this preliminary result of $M^{}_H$
on the SM vacuum stability by using the two-loop
renormalization-group equations (RGEs), and arrive at the cutoff
scale $\Lambda^{}_{\rm VS} \sim 4 \times 10^{12}$ GeV (for $M^{}_H =
125 ~{\rm GeV}$, $M^{}_t = 172.9~{\rm GeV}$ and $\alpha^{}_s(M^{}_Z)
= 0.1184$) where the absolute stability of the SM vacuum is lost and
some kind of new physics might take effect. We update the values of
running lepton and quark masses at some typical energy scales,
including the ones characterized by $M^{}_H$, 1 TeV and
$\Lambda^{}_{\rm VS}$, with the help of the two-loop RGEs. The
branching ratios of some important two-body Higgs decay modes, such
as $H \to b\bar{b}$, $H \to \tau^+ \tau^-$, $H\to \gamma\gamma$,
$H\to W^+ W^-$ and $H \to Z Z$, are also recalculated by inputting
the values of relevant particle masses at $M^{}_H$.
\end{abstract}


\section{Introduction}

The Higgs mechanism \cite{Higgs} is responsible for the spontaneous
${\rm SU(2)}_{\rm L} \otimes {\rm U(1)}_{\rm Y} \to {\rm U(1)}_{\rm
em}$ gauge symmetry breaking in the standard model (SM) of
electroweak interactions \cite{SM}, but the Higgs boson itself left
no sort of trace in all the previous high-energy collider
experiments. The main goal of the Large Hadron Collider (LHC) at
CERN is just to discover this elusive particle, which allows other
particles (except the photon and gluons) to gain finite masses.
Combined with the indirect bounds obtained from the electroweak
precision measurements, the recent data of the ATLAS and CMS
experiments lead us to a rather narrow range of the Higgs mass:
$114~{\rm GeV} \lesssim M^{}_H \lesssim 141~{\rm GeV}$ \cite{114}.
In particular, both collaborations have reported
their latest results
\begin{equation}
M^{}_H \simeq \left\{ \begin{array}{l} \left(116 \cdots 131 \right) ~{\rm
GeV} ~~~~({\rm ATLAS} ~\cite{ATLAS}) \; , \\
\left(115 \cdots 127 \right) ~{\rm GeV} ~~~~({\rm CMS} ~\cite{CMS})
\; , \end{array} \right.
\end{equation}
at the $95\%$ confidence level. The ATLAS Collaboration has also
found a preliminary hint of $M^{}_H \simeq
126~{\rm GeV}$ with the $3.6\sigma$ local significance in
$H \to \gamma \gamma$ ($2.8\sigma$), $H \to Z Z^* \to 4
l$ ($2.1\sigma$) and $H \to W W^* \to 2l2\nu$ ($1.4\sigma$)
decay modes \cite{ATLAS}; and the CMS Collaboration has observed an
excess compatible with $M^{}_H \lesssim 124~{\rm GeV}$ with the
$2.6\sigma$ local significance \cite{CMS}. These interesting results
point to a preferred and narrower range
$M^{}_H \simeq (124 \cdots 126) ~{\rm GeV}$ for the SM Higgs
boson. We are therefore confident that an unambiguous discovery
of the Higgs boson will soon come true at the LHC.

Observing the Higgs boson and measuring its mass and other
properties may help us solve several fundamental problems in
elementary particle physics. Here we mention three of them for
example.
\begin{itemize}
\item     The Higgs mass theoretically suffers significant
radiative corrections, and hence new symmetries and (or) new
particles should be introduced to stabilize the electroweak scale
$\Lambda^{}_{\rm EW} \sim 10^2~{\rm GeV}$ \cite{Hierarchy}. A
solution to this gauge hierarchy problem calls for new physics
beyond the SM, such as supersymmetries \cite{SUSY} or extra spatial
dimensions \cite{extra}.

\item     The Higgs boson is indispensable to the
Yukawa interactions of three-family fermions which makes weak CP
violation possible in the SM or its simple extensions \cite{KM}. To
some extent, the existence of a Higgs boson may also support the
Peccei-Quinn mechanism as an appealing solution to the strong CP
problem \cite{PQ}.

\item     With the help of the Higgs field, one may write
out the unique dimension-five operator $\ell \ell H H$ in an
effective field theory \cite{D5} or implement the seesaw mechanism
in a renormalizable quantum field theory \cite{SS1} to generate
finite but tiny neutrino masses.
\end{itemize}
Therefore, the highest priority of the LHC experiment is to pin down
the Higgs boson and its quantum numbers. We are approaching
a success in this connection.

Motivated by the encouraging ATLAS and CMS results, we aim to
examine the impacts of $M^{}_H \simeq (124 \cdots 126)~{\rm GeV}$ on
the vacuum stability of the SM, the running behaviors of fermion
masses and the branching ratios of the Higgs decays. The point is
that a relatively small value of $M^{}_H$ is likely to cause the
vacuum instability unless new physics takes effect at a proper
cutoff scale \cite{vacuum}. Given $M^{}_H \simeq 125$ GeV as
indicated by the latest LHC data, it is timely to determine the
energy scale at which the effective quartic Higgs coupling
$\tilde{\lambda} (\mu)$ runs to zero. We find that this cutoff scale
is around $\Lambda^{}_{\rm VS} \sim 4\times 10^{12}$ GeV, which
presumably signifies the end of the gauge desert and the beginning
of a new physics oasis. Taking account of the allowed range of
$M^{}_H$ and the updated values of other SM parameters, we
recalculate the running fermion masses at some typical energy scales
up to $\Lambda^{}_{\rm VS}$ by means of the renormalization-group
equations (RGEs). Such an exercise makes sense because a
sufficiently large value of $M^{}_H$ (e.g., $M^{}_H \simeq 140$ GeV)
was assumed in the previous works and hence the potential vacuum
stability problem did not show up \cite{xzz}. As a by-product, the
branching ratios of some important two-body Higgs decay modes in the
SM, such as $H \to b\bar{b}$, $H \to \tau^+ \tau^-$, $H\to
\gamma\gamma$, $H\to W^+W^-$ and $H \to Z Z$, are also recalculated
by using the new values of relevant particle masses obtained at $\mu
\sim M^{}_H$.

\section{The Higgs mass and vacuum stability}

First of all, let us briefly review the vacuum stability issue in
the SM with a relatively light Higgs boson. In order to find out the
true vacuum state and analyze its stability, one should calculate
the effective scalar potential by taking account of radiative
corrections and RGE improvements of the relevant parameters
\cite{vacuum,Hunter}. It has been shown that the $L$-loop scalar
potential improved with $(L+1)$-loop RGEs actually includes all the
$L$th-to-leading logarithm contributions \cite{Bando}. At the
one-loop level, the effective scalar potential in the 't
Hooft-Landau gauge and in the $\overline{\rm MS}$ renormalization
scheme can be written as \cite{potential}
\begin{eqnarray}
V^{}_{\rm eff} \left[ \phi(t) \right] &=& -\frac{1}{2} m^2(t)
\phi^2(t) + \frac{1}{4} \lambda(t) \phi^4(t) + \frac{3}{64\pi^2}
\left\{ 2 m^4_W\left[ \phi(t) \right] \left[ \ln
\left(\frac{m^2_W\left[ \phi(t) \right]}{\mu^2(t)}\right) -
\frac{5}{6}\right] \right. \nonumber \\
&~& \left. + m^4_Z\left[ \phi(t) \right] \left[ \ln \left(
\frac{m^2_Z\left[ \phi(t) \right]}{\mu^2(t)} \right) -
\frac{5}{6}\right]- 4 m^4_t\left[ \phi(t) \right] \left[ \ln
\left(\frac{m^2_t\left[ \phi(t) \right]}{\mu^2(t)} \right) -
\frac{3}{2}\right]\right\} \;,
\end{eqnarray}
where the contributions from the Goldstone and Higgs bosons have
been safely neglected, and $m^2_W\left[ \phi(t) \right] \equiv
g^2(t) \phi^2(t)/4$, $m^2_Z\left[ \phi(t) \right] \equiv [g^2(t) +
{g^\prime}^2(t)] \phi^2(t)/4$ and $m^2_t\left[ \phi(t) \right]
\equiv y^2_t(t) \phi^2(t)/2$ have been defined. Note that the scale
dependence of all the dimensionless couplings $(g(t), g^\prime(t),
\lambda(t), y^{}_t(t))$, the mass parameter $m^2(t)$ and the Higgs
field $\phi(t)$ has been explicitly indicated through the
renormalization scale $\mu(t) \equiv M^{}_Z e^t$ or equivalently the
running parameter $t = \ln \left[\mu(t)/M^{}_Z\right]$. The $\beta$
functions for the dimensionless couplings $(g(t), g^\prime(t),
\lambda(t), y^{}_t(t))$ and the $\gamma$ functions for
$(m^2(t),\phi(t))$ at the two-loop order can be found in Refs.
\cite{potential,RGEs}.

Due to the experimental observations, the scalar potential
$V^{}_{\rm eff}$ must develop a realistic minimum at the electroweak
scale, corresponding to the SM vacuum. Whether the SM vacuum is
stable or not depends on the behavior of $V^{}_{\rm eff}$ in the
large-field limit, i.e., $\phi(t) \gg M^{}_Z$. More explicitly, one
can find out the extrema $\phi^{}_{\rm ex}(t)$ of the scalar
potential via
\begin{eqnarray}
\left. \frac{\partial V^{}_{\rm eff}[\phi(t)]}{\partial \phi(t)}
\right|^{}_{\phi(t) = \phi^{}_{\rm ex}(t)} = 0 \; .
\end{eqnarray}
At the weak scale $\mu(t^{}_Z) = M^{}_Z$, we should impose the
boundary condition $\phi^{}_{\rm ex}(t^{}_Z) = v \approx 246~{\rm
GeV}$, which is the vacuum expectation value of the Higgs field. At
the large-field values, the scalar potential is dominated by the
quartic coupling term and the extrema $\phi^{}_{\rm ex}(t)$ can be
evaluated at the renormalization scale $\mu(t) = \phi^{}_{\rm
ex}(t)$ from Eqs. (2) and (3) as $\phi^2_{\rm ex} =
m^2/\tilde{\lambda}$, where the effective quartic coupling
$\tilde{\lambda}$ is defined as
\begin{eqnarray}
\tilde{\lambda} = \lambda &-& \frac{3}{32\pi^2}
\left\{\frac{1}{8}({g^\prime}^2 + g^2)^2 \left[\frac{1}{3} - \ln
\left(\frac{{g^\prime}^2 + g^2}{4}\right)\right]  \right. \nonumber
\\
 &+& \left. 2 y^4_t \left[ \ln \left(\frac{y^2_t}{2}\right) -
1\right] + \frac{1}{4} g^4 \left[\frac{1}{3} -
\ln\left(\frac{g^2}{4}\right)\right]\right\} \; .
\end{eqnarray}
Now it is clear that $V^{}_{\rm eff} \approx \tilde{\lambda}
\phi^4/4$ will develop a minimum much deeper than the realistic
minimum if the effective coupling $\tilde{\lambda}$ becomes negative
\cite{CEQ,Altarelli,Giudice,triviality,Hambye}. To maintain the
absolute stability of the SM vacuum, new physics should come into
play below or at the energy scale $\Lambda^{}_{\rm VS}$ where the
effective coupling $\tilde{\lambda}$ vanishes, i.e.,
$\tilde{\lambda}(\Lambda^{}_{\rm VS}) = 0$. One can derive a lower
mass bound on the Higgs boson by requiring that the SM vacuum is
absolutely stable up to a possible grand-unified-theory (GUT) scale or the
Planck scale \cite{vacuum,CEQ,Altarelli,Giudice,triviality,Hambye}.

In view of the allowed range of the Higgs mass, we may conversely
implement the vacuum stability argument to determine the energy
scale $\Lambda^{}_{\rm VS}$ at which new physics should take effect.
Our strategy is as follows. First, we have to specify the matching
conditions relating the quartic coupling $\lambda$ to the Higgs mass
$M^{}_H$, as well as the top-quark Yukawa coupling $y^{}_t$ to the
top-quark pole mass $M^{}_t$. Although the complete effective
potential $V^{}_{\rm eff}$ must be scale-independent, the one with
one-loop approximation is not. The solution is to find an optimal
scale $\mu^* = \mu(t^*)$ for which the effective potential has the
least scale-dependence, as shown in Ref. \cite{CEQ}, where one can
observe that $\mu^* = M^{}_t$ is a reasonable choice. Therefore, we
choose the matching conditions for $\lambda$ and $y^{}_t$ at $\mu^*
= M^{}_t$:
\begin{eqnarray}
\lambda(M^{}_t) &=& \frac{M^2_H}{2v^2} \left[ 1 +
\delta^{}_H(M^{}_t)\right] \;, \nonumber \\
y^{}_t(M^{}_t) &=& \frac{\sqrt{2} M^{}_t}{v} \left[ 1 +
\delta^{}_t(M^{}_t)\right] \; ,
\end{eqnarray}
where the correction terms $\delta^{}_H(M^{}_t)$ and
$\delta^{}_t(M^{}_t)$ have been given in
Ref.\cite{Hambye,Sirlin,Kniehl}. The values of the other input
parameters are taken from Ref. \cite{PDG} and will be specified in
Sec. III when we turn to the running fermion masses. Second, we run
$\lambda(\mu)$ to a much higher energy scale by solving the complete
two-loop RGEs. Third, the cutoff scale $\Lambda^{}_{\rm VS}$ can be
identified with the solution to $\tilde{\lambda}(\Lambda^{}_{\rm
VS}) = 0$, where $\tilde{\lambda}$ and $\lambda$ are related via Eq.
(4). Note that the cutoff scale $\Lambda^{}_{\rm VS}$ determined by
$\tilde{\lambda}(\Lambda^{}_{\rm VS}) = 0$ could be an order of
magnitude larger than the one by $\lambda(\Lambda^{}_{\rm VS}) = 0$,
which has not taken account of the one-loop radiative corrections to
the scalar potential \cite{CEQ,Giudice}.

Our numerical result for the correlation between the Higgs mass and
the energy scale is shown in FIG. 1. Some comments are in order.
\begin{enumerate}
\item If $M^{}_H \gtrsim 129~{\rm GeV}$ holds, the vacuum stability
can be guaranteed even around a possible GUT scale
(e.g., $10^{16}~{\rm GeV}$) or the Planck scale $\Lambda^{}_{\rm Pl}
\sim 10^{19}~{\rm GeV}$\cite{Lindner}. The cutoff scale
$\Lambda^{}_{\rm VS}$ increases as the Higgs mass $M^{}_H$
increases, but this observation is sensitively dependent on the
value of the top-quark pole mass $M^{}_t$.

\item Given $M^{}_H \simeq 125~{\rm GeV}$, some kind of
new physics should come out around $\Lambda^{}_{\rm VS} \sim
10^{12}~{\rm GeV}$ to stabilize the SM vacuum
\footnote{Since the cutoff scale depends sensitively on the Higgs
mass in the range $[120~{\rm GeV}, 130~{\rm GeV}]$, as shown in FIG.
1, one has to take care of experimental errors from the top quark
mass $M^{}_t$ and the strong coupling $\alpha^{}_s(M^{}_Z)$, as well
as the theoretical uncertainties involved in the two-loop RGEs and
one-loop matching conditions \cite{Giudice}. For instance, the SM
vacuum for $M^{}_H \simeq 125~{\rm GeV}$ could even be stable up to
the Planck scale $\Lambda^{}_{\rm Pl} \simeq 10^{19}~{\rm GeV}$ if
the relevant uncertainties are included.}.
For example, it is interesting to notice that the
canonical seesaw mechanism for neutrino mass generation is expected
to work around this cutoff scale. In such a seesaw model the heavy
Majorana neutrinos could have masses of ${\cal O}(10^{12})~{\rm
GeV}$, so that the leptogenesis mechanism \cite{FY} may work well to
account for the observed matter-antimatter asymmetry of the
Universe.
\end{enumerate}
Although the existence of a cutoff scale is robust for the SM with a
relatively light Higgs boson, it remains unclear what kind of new
physics could take effect over there. In any event, if the new
physics responsible for the vacuum stability could also offer a
solution to the flavor puzzles of leptons and quarks (especially the
origin of tiny neutrino masses), the running fermion masses at the
cutoff scale $\Lambda^{}_{\rm VS}$ will be very helpful for model
building. We shall focus on this issue in the following section.

\section{Running lepton and quark masses}

A systematic analysis of the RGE running masses of leptons and quarks
has been done in Ref. \cite{xzz}, where $M^{}_H \simeq 140~{\rm GeV}$ has
typically been taken just for illustration.
As discussed above, such a value of the Higgs mass makes the situation
simple because it does not give rise to the vacuum instability problem
in the SM. Here we want to update the running fermion masses for two
good reasons: (a) the latest ATLAS and CMS data point to
$M^{}_H \simeq (124 \cdots 126)~{\rm GeV}$,
and hence the issue of vacuum stability should be taken seriously;
(b) the values of some of the input parameters adopted in Ref.
\cite{xzz} have more or less changed in the past few years, and thus
an update of the analysis is necessary. Before doing a detailed RGE
analysis of fermion masses, let us summarize the input
parameters and outline our calculational strategy.
\begin{itemize}
\item Six quark masses are
$m^{}_u(2~{\rm GeV}) = (1.7 \cdots 3.1)~{\rm MeV}$, $m^{}_d(2~{\rm
GeV}) = (4.1 \cdots 5.7)~{\rm MeV}$, $m^{}_s(2~{\rm GeV}) = (80
\cdots 130)~{\rm MeV}$, $m^{}_c (m^{}_c) = 1.29^{+0.05}_{-0.11}~{\rm
GeV}$, $m^{}_b (m^{}_b) = 4.19^{+0.08}_{-0.16}~{\rm GeV}$ and
$M^{}_t = 172.9^{+1.1}_{-1.1}~{\rm GeV}$ \cite{PDG}, where $M^{}_t$
represents the pole mass of the top quark extracted from the direct
measurements. In addition, the pole masses of three charged leptons
are given by $M^{}_e = (0.510998910\pm 0.000000013)~{\rm MeV}$,
$M^{}_\mu = (105.658367 \pm 0.000004)~{\rm MeV}$ and $M^{}_\tau =
(1776.82 \pm 0.16)~{\rm MeV}$ \cite{PDG}. Following the same
approach as the one described in Ref. \cite{xzz}, we can calculate
the running masses of charged leptons and quarks at some typical
energy scales in the SM, including $\mu = M^{}_W$, $M^{}_Z$,
$M^{}_H$, 1 TeV and $\Lambda^{}_{\rm VS}$.

\item The strong and electromagnetic fine-structure constants at
$M^{}_Z$ are $\alpha^{}_s(M_Z) = 0.1184 \pm 0.0007$ and
$\alpha(M_Z)^{-1} = 127.916\pm 0.015$, and the weak mixing angle is
$\sin^2 \theta^{}_W (M^{}_Z) = 0.231 16 \pm 0.00013$ \cite{PDG}.
With the help of these input parameters, one may determine the gauge
coupling constants $g^2_s = 4\pi \alpha^{}_s$, $g^2 = 4\pi
\alpha/\sin^2 \theta^{}_W$ and $g^\prime = g \tan \theta^{}_W$ at
the energy scale $\mu = M^{}_Z$.

\item The four parameters of quark flavor mixing and CP violation
in the modified Wolfenstein parametrization are $\lambda = 0.2253
\pm 0.0007$, $A = 0.808^{+0.022}_{-0.015}$, $\bar \rho =
0.132^{+0.022}_{-0.014}$ and $\bar\eta = 0.341 \pm 0.013$
\cite{PDG}. These values, together with the values of quark masses,
allow us to reconstruct the quark Yukawa coupling matrices
$Y^{}_{\rm u}$ and $Y^{}_{\rm d}$ at the electroweak scale. The RGEs
of $Y^{}_{\rm u}$ and $Y^{}_{\rm d}$ can therefore help us to run
the quark masses and flavor mixing parameters to a much higher
energy scale.

\item The allowed ranges of three lepton flavor mixing angles are
$30.6^\circ \leq \theta^{}_{12} \leq 36.8^\circ$, $35.7^\circ \leq
\theta^{}_{23} \leq 53.1^\circ$ and $1.8^\circ \leq \theta^{}_{13}
\leq 12.1^\circ$ \cite{Fogli}, and the allowed ranges of two
neutrino mass-squared differences are $6.99\times 10^{-5}~{\rm eV}^2
\leq \delta m^2 \leq 8.18 \times 10^{-5}~{\rm eV}^2$ and $2.06\times
10^{-3}~{\rm eV}^2 \leq |\Delta m^2| \leq 2.67 \times 10^{-3}~{\rm
eV}^2$ \cite{Fogli}. For simplicity, we only take the best-fit
values $\theta^{}_{12} = 33.6^\circ$, $\theta^{}_{23} = 40.4^\circ$
and $\theta^{}_{13} = 8.3^\circ$ together with $\delta m^2 =
7.58\times 10^{-5}~{\rm eV}^2$ and $|\Delta m^2| = 2.35\times
10^{-3}~{\rm eV}^2$ as the inputs at $M^{}_Z$ in our numerical
calculations. In particular, the value of $\theta^{}_{13}$ taken
above is essentially consistent with the latest Daya Bay \cite{DYB}
and RENO \cite{RENO} results.
The unknown CP-violating phases in the lepton sector
are all assumed to be zero. In view of the fact that the absolute
neutrino mass scale is also unknown, we shall only consider the
normal mass hierarchy with $m^{}_1 = 0.001~{\rm eV}$ and $m^{}_1 <
m^{}_2 \ll m^{}_3$ at $M^{}_Z$ for illustration. For the same
reason, only the one-loop RGE for neutrino masses is considered. It is
then possible to reconstruct the charged-lepton Yukawa coupling
matrix $Y^{}_l$ and the effective neutrino coupling matrix $\kappa$
at $M^{}_Z$ from the given lepton masses and flavor mixing
parameters \cite{XZ}.
\end{itemize}
For a complete list of the RGEs to be used in our numerical
analysis, we refer the reader to Ref. \cite{xzz} and references
therein.

TABLES I and II summarize our numerical results for the running
quark and charged-lepton masses at some typical energy scales,
respectively. Different from the previous works, here the scales
characterized by the Higgs mass $M^{}_H$ and the vacuum stability
cutoff $\Lambda^{}_{\rm VS}$ are taken into account for the first
time. The values of fermion masses at $M^{}_H$ will be used to
calculate the branching ratios of some important Higgs decay modes
in Sec. IV, and those at $\Lambda^{}_{\rm VS}$ are expected to be
very useful for building possible flavor models beyond the SM.

In studying the running behaviors of twelve fermion masses above
$M^{}_Z$, we have used the inputs at $M^{}_Z$ to numerically
solve the RGEs of the Yukawa coupling matrices $Y^{}_{\rm u}$,
$Y^{}_{\rm d}$, $Y^{}_l$ and the effective neutrino coupling matrix
$\kappa$ as well as the two-loop RGEs of the quartic Higgs coupling
$\lambda(\mu)$ and gauge couplings at $\mu \geq M^{}_Z$. After
$Y^{}_{\rm u}$, $Y^{}_{\rm d}$, $Y^{}_l$ and $\kappa$ are
diagonalized, one can obtain the running quark masses $m^{}_q(\mu) =
y^{}_q(\mu) \ v/\sqrt{2}$ (for $q=u, c, t$ and $d, s, b$), the
running charged-lepton masses $m^{}_l (\mu) = y^{}_l (\mu) \
v/\sqrt{2}$ (for $l = e, \mu, \tau$) and the running neutrino masses
$m^{}_i (\mu) = \kappa^{}_i(\mu) \ v^2/2$ (for $i=1,2,3$). The
corresponding quark and lepton flavor mixing parameters can
simultaneously be achieved. For simplicity, let us define
$R^{}_f(\mu) \equiv m^{}_f(\mu)/m^{}_f(M^{}_Z)$, where the subscript
$f$ runs over the mass-eigenstate indices of six quarks and six
leptons. We find that $R^{}_u(\mu) \approx R^{}_d(\mu) \approx
R^{}_s(\mu) \approx R^{}_c(\mu) \approx R^{}_e(\mu) \approx
R^{}_\mu(\mu) \approx 1$ holds to a good degree of accuracy, if
$\mu$ is below the cutoff scale $\Lambda^{}_{\rm VS}$. So we only
plot the numerical results of $R^{}_t(\mu)$, $R^{}_b(\mu)$ and
$R^{}_\tau(\mu)$ in FIG. 2. The ratios $R^{}_i(\mu)$ for three
neutrino masses are shown in FIG. 3. Some comments are in order.
\begin{enumerate}

\item The mass ratios $R^{}_f(\mu)$ are not very sensitive to
the quartic Higgs coupling $\lambda(\mu)$ or equivalently the Higgs
mass $M^{}_H$, simply because the latter enters the RGEs of
fermion masses only at the two-loop level. As observed in
Ref. \cite{xzz}, there exists a maximum for the charged-lepton
masses around $\mu \sim 10^6~{\rm GeV}$, while the quark masses
monotonously decrease as the energy scale increases. Taking account
of the vacuum instability problem discussed in Sec. II, we argue
that the evolution of fermion masses above the cutoff scale
$\Lambda^{}_{\rm VS}$ might not be meaningful anymore. We expect
that some kind of new physics should take effect around
$\Lambda^{}_{\rm VS}$ and thus modify the RGEs of the SM.

\item In most cases the running behaviors of three
neutrino masses are neither sensitive to their absolute values nor
sensitive to their mass hierarchies in the SM \cite{Antusch}. Only
when three neutrino masses are assumed to be nearly degenerate,
the RGE running effects of neutrino mass and mixing parameters are
possible to
be significant. But the dependence of $m^{}_i(\mu)$ on the quartic
Higgs coupling $\lambda(\mu)$ or the Higgs mass $M^{}_H$ is quite
evident, simply because the effective neutrino coupling matrix
$\kappa$ receives the one-loop corrections from the quartic Higgs
interaction \cite{xzz,Antusch}.
\end{enumerate}
For simplicity, we skip the numerical illustration of the running
behaviors of quark and lepton flavor mixing parameters in this
paper.

\section{Branching ratios of the Higgs decays}

The present ATLAS and CMS experiments are mainly sensitive to the
Higgs boson via its decay channels $H \to \gamma \gamma$, $H \to b
\bar{b}$, $H \to \tau^+ \tau^-$, $H \to W^+ W^- \ (2l2\nu)$ and $H
\to Z Z \ (4 l, 2l 2\nu, 2l2q, 2l 2\tau)$, where $l = e$ or $\mu$
and $\nu$ denotes the neutrinos of any flavors \cite{114}. Which
channel is dominant depends crucially on the Higgs mass. If $M^{}_H
\lesssim 135~{\rm GeV}$ holds,the decay mode $H \to b \bar{b}$ is
expected to have the largest branching ratio; and if the Higgs mass
is slightly heavier, the decay mode $H \to W^+ W^-$ will surpass the
others \cite{Hunter}.

We first consider the leptonic $H \to l^+ l^-$ decays, where $l$
runs over $e$, $\mu$ or $\tau$. Including the one-loop electroweak
corrections, the decay width of $H \to l^+ l^-$ is given by
\cite{EW}
\begin{equation}
\Gamma^{}_l = \frac{G^{}_{\rm F} M^{}_H}{4\sqrt{2}\pi} M^2_l \left(1
- \frac{4M^2_l}{M^2_H} \right)^{3/2} \left(1 + \delta^{}_{\rm QED} +
\delta^{}_{\rm W} \right) \; ,
\end{equation}
where $G^{}_{\rm F}$ is the Fermi constant, $\delta^{}_{\rm QED} =
9\alpha \left[3 - 2\ln
\left(M^2_H/M^2_l\right)\right]/\left(12\pi\right)$, and
\begin{equation}
\delta^{}_{\rm W} = \frac{G^{}_{\rm F}}{8\sqrt{2}\pi^2}
\left\{7M^2_t + M^2_W \left( \frac{3}{\sin^2\theta^{}_W} \ln
\cos^2\theta^{}_W - 5\right) -
M^2_Z\left[3\left(1-4\sin^2\theta^{}_W\right)^2 -
\frac{1}{2}\right]\right\} \; .
\end{equation}
Note that the large logarithmic term $\ln(M^2_H/M^2_l)$ in
$\delta^{}_{\rm QED}$ can be absorbed in the running mass of $l$ at
the scale of $M^{}_H$, which has been given in TABLE II.

Now we turn to the $H \to q \bar{q}$ decays, where $q$ runs over
$u$, $d$, $s$, $c$ or $b$ for the Higgs mass to lie in the range
$114~{\rm GeV} \lesssim M^{}_H \lesssim 141~{\rm GeV}$. Since the
decay rates of $H \to u\bar{u}$, $d\bar{d}$ and $s\bar{s}$ are
highly suppressed by the corresponding quark masses, we are mainly
interested in the decay rates of $H \to c\bar{c}$ and $H\to b
\bar{b}$. Up to the three-loop QCD corrections \cite{QCD3},
\begin{equation}
\Gamma^{}_q = \frac{3G^{}_{\rm F} M^{}_H}{4\sqrt{2}\pi}
m^2_q(M^{}_H) \left(\delta^{}_{\rm QCD} + \delta^{}_t\right) \; ,
\end{equation}
where
\begin{equation}
\delta^{}_{\rm QCD} = 1 + 5.67
\left(\frac{\alpha^{}_s(M^{}_H)}{\pi}\right) + 29.14
\left(\frac{\alpha^{}_s(M^{}_H)}{\pi}\right)^2 + 41.77
\left(\frac{\alpha^{}_s(M^{}_H)}{\pi}\right)^3 \; ;
\end{equation}
and
\begin{equation}
\delta^{}_t = \left(\frac{\alpha^{}_s(M^{}_H)}{\pi}\right)^2
\left[1.57 - \frac{2}{3} \ln \left(\frac{M^2_H}{M^2_t}\right) +
\frac{1}{9} \ln^2 \left(\frac{m^2_q(M^{}_H)}{M^2_H}\right)\right]
\;.
\end{equation}
Note that the running quark masses $m^{}_q(M^{}_H)$ and the strong
coupling constant $\alpha^{}_s(M^{}_H)$ at the scale of $M^{}_H$ are
useful here to absorb the large logarithmic terms.

A detailed discussion about the two-body decay modes $H \to
\gamma\gamma$, $H \to W^+W^-$, $H\to ZZ$, $H \to gg$ and $H \to
t\bar{t}$ can be found in Ref. \cite{Spira}. For simplicity, here we
do not elaborate the relevant analytical results but do a numerical
recalculation based on the updated particle masses at $M^{}_H$. In
order to compute the branching ratios of the above decay channels,
we implement the latest version of the program HDECAY \cite{HDECAY}
and update the input parameters according to our TABLES I and II
together with Ref. \cite{PDG}. Some comments are in order.
\begin{itemize}
\item The pole masses of the charged leptons (i.e., $M^{}_l$),
instead of the running masses $m^{}_l(M^{}_H)$, have been used as
the input parameters in the program HDECAY. This treatment is
consistent with the formula of $\Gamma^{}_l$ given in Eq. (6). If
one chooses to use the running masses $m^{}_l(M^{}_H)$ in the numerical
calculation, then the correction terms in Eq. (6) should take
different forms.

\item The one-loop pole masses of $c$ and $b$ quarks have been
used as the input parameters in the program HDECAY, because they
must be consistent with the corresponding parton distribution
function when the production of the Higgs boson in a hadron collider
(e.g., the LHC) is taken into account \cite{Denner}. In our
calculations we start from the values of $m^{}_c(m^{}_c)$ and
$m^{}_b(m^{}_b)$ \cite{PDG} and then evaluate the pole masses
$M^{}_c$ and $M^{}_b$ as precisely as possible by using the relevant
four-loop RGEs and three-loop matching conditions \cite{xzz}. Hence
we obtain the pole masses $M^{}_c = 1.84~{\rm GeV}$ and $M^{}_b =
4.92~{\rm GeV}$, as given in TABLE I.
\end{itemize}
Our numerical results for the branching ratios of $H \to b \bar{b}$,
$c \bar{c}$ and $\tau^+\tau^-$ decays are shown in FIG. 4, where the
branching ratios of $H \to \gamma \gamma$, $g g$, $W^+ W^-$, $Z Z$
and $Z \gamma$ decays are also plotted for a comparison.
These important two-body decay channels will help discover the
Higgs boson and pin down its mass in the near future. The
branching ratios of $H \to s \bar{s}$ and $H \to \mu^+\mu^-$ are of
${\cal O}(10^{-4})$ in the range of $M^{}_H \in [110~{\rm GeV},
150~{\rm GeV}]$, and thus they have been neglected from FIG. 4.

\section{Summary}

In view of the recent results from the ATLAS and CMS experiments
which hint at the existence of the Higgs boson, we have examined the
impact of the Higg mass on vacuum stability in the SM by means of
the two-loop RGEs. We find that $M^{}_H \simeq 125$ GeV leads us to
an interesting cutoff scale $\Lambda^{}_{\rm VS} \sim 10^{12}$ GeV,
as required by the vacuum stability. Some kind of new physics are
therefore expected to take effect around $\Lambda^{}_{\rm VS}$. In
other words, $\Lambda^{}_{\rm VS}$ characterizes the end of the
gauge desert and the beginning of a new physics oasis.

We have argued that possible new physics responsible for the vacuum
stability of the SM might also be able to help solve the flavor
puzzles. Hence we have recalculated the running fermion masses up to
the cutoff scale $\Lambda^{}_{\rm VS}$ by inputting the allowed range
of $M^{}_H$ and the updated values of other SM parameters into the
full set of the two-loop RGEs for the quartic Higgs coupling, the
Yukawa couplings and the gauge couplings. In particular, the values
of lepton and quark masses at $\mu = M^{}_H$ and $\Lambda^{}_{\rm
VS}$ are obtained for the first time. As a by-product, the branching
ratios of some important two-body Higgs decay modes in the SM, such
as $H \to b\bar{b}$, $H\to \tau^+ \tau^-$, $H\to \gamma\gamma$,
$H\to W^+W^-$ and $H \to ZZ$, have been recalculated with the help
of the new values of relevant particle masses obtained at $M^{}_H$.
Our numerical results should be very useful for model building and
flavor physics.

We reiterate that an unambiguous discovery of the Higgs boson at the
LHC in the near future will pave the way for us to confirm the
Yukawa interactions between the Higgs field and fermion fields. That
will be a crucial step towards understanding the origin of fermion
masses, flavor mixing and CP violation either within or beyond the
SM. This point is especially true for testing the seesaw mechanisms,
which attribute the tiny masses of three known neutrinos to the
presence of some unknown heavy degrees of freedom via the Yukawa
interactions. We believe that a new era of flavor physics is coming
to the surface.

\vspace{0.5cm}

The authors are indebted to Peter Zerwas for valuable comments and
suggestions. This work was supported in part by the National Natural
Science Foundation of China under grant No. 11135009 and
by the Ministry of Science and Technology of China under grant No.
2009CB825207 (Z.Z.X.), by
the ERC under the Starting Grant MANITOP and by the DFG in the
Transregio 27 ``Neutrinos and Beyond'' (H.Z.) and by the Alexander
von Humboldt Foundation (S.Z.).


\newpage


\begin{table}[]
\caption{Running quark masses at some typical energy scales in the
SM, including the Higgs mass $M^{}_H \simeq 125~{\rm GeV}$ and the
corresponding cutoff scale $\Lambda^{}_{\rm VS} \simeq 4\times
10^{12}~{\rm GeV}$. Note that the values of the pole masses $M^{}_q$
and running masses $m^{}_q(M^{}_q)$ themselves, rather than the
running masses $m^{}_q(\mu)$ at these mass scales, are given in the
last two rows for comparison. But the pole masses of three light
quarks are not listed, simply because the perturbative QCD
calculation is not reliable in that energy region.}
\begin{center}
\begin{tabular}{c|c|c|c|c|c|c}
$\mu$ & $m^{}_u(\mu)~ ({\rm MeV})$ &  $m^{}_d(\mu)~ ({\rm MeV})$ &
$m^{}_s(\mu)~ ({\rm MeV})$ & $m^{}_c(\mu)~ ({\rm GeV})$
& $m^{}_b(\mu)~ ({\rm GeV})$ & $m^{}_t(\mu)~ ({\rm GeV})$ \\
\hline
$m^{}_c(m^{}_c)$ & $2.79^{+0.83}_{-0.82}$ & $5.69^{+0.96}_{-0.95}$ &
$116^{+36}_{-24}$ & $1.29^{+0.05}_{-0.11}$ & $5.95^{+0.37}_{-0.15}$
& $385.7^{+8.1}_{-7.8}$\\
\hline
$2 ~{\rm GeV}$ & $2.4^{+0.7}_{-0.7}$ & $4.9 \pm 0.8$  &
$100^{+30}_{-20}$
& $1.11^{+0.07}_{-0.14}$ & $5.06^{+0.29}_{-0.11}$ & $322.2^{+5.0}_{-4.9}$ \\
\hline
$m^{}_b(m^{}_b)$ & $2.02^{+0.60}_{-0.60}$ & $4.12^{+0.69}_{-0.68}$ &
$84^{+26}_{-17}$ & $0.934^{+0.058}_{-0.120}$ &
$4.19^{+0.18}_{-0.16}$
& $261.8^{+3.0}_{-2.9}$ \\
\hline
$M^{}_W$ & $1.39^{+0.42}_{-0.41}$ & $2.85^{+0.49}_{-0.48}$ &
$58^{+18}_{-12}$
& $0.645^{+0.043}_{-0.085} $ & $2.90^{+0.16}_{-0.06}$ & $174.2 \pm 1.2$  \\
\hline
$M^{}_Z$ & $1.38^{+0.42}_{-0.41}$ & $2.82 \pm 0.48$ &
$57^{+18}_{-12}$
& $0.638^{+0.043}_{-0.084}$ & $2.86^{+0.16}_{-0.06}$ & $172.1 \pm 1.2 $ \\
\hline
$M^{}_H$ & $1.34^{+0.40}_{-0.40}$ & $2.74^{+0.47}_{-0.47}$ &
$56^{+17}_{-12}$ & $0.621^{+0.041}_{-0.082}$ & $2.79^{+0.15}_{-0.06}$ &
$167.0^{+1.2}_{-1.2}$ \\
\hline
$m^{}_t(m^{}_t)$ & $1.31^{+0.40}_{-0.39}$ & $2.68 \pm 0.46 $ &
$55^{+17}_{-11}$ & $0.608^{+0.041}_{-0.080}$ &
$2.73^{+0.15}_{-0.06}$
& $163.3 \pm 1.1$ \\
\hline
$1~{\rm TeV}$ & $1.17 \pm 0.35$ & $2.40^{+0.42}_{-0.41} $ &
$49^{+15}_{-10}$ & $0.543^{+0.037}_{-0.072}$ &
$2.41^{+0.14}_{-0.05}$
& $148.1 \pm 1.3$ \\
\hline
$\Lambda^{}_{\rm VS}$ & $0.61^{+0.19}_{-0.18}$ & $1.27 \pm 0.22$ &
$26^{+8}_{-5}$ & $0.281^{+0.02}_{-0.04}$ & $1.16^{+0.07}_{-0.02}$
& $82.6 \pm 1.4$\\
\hline \hline
$M^{}_q$ & $-$ & $-$ & $-$ & $1.84^{+0.07}_{-0.13}$
& $4.92^{+0.21}_{-0.08}$ & $172.9 \pm 1.1$ \\
\hline
$m^{}_q(M^{}_q)$ & $-$ & $-$ & $-$ & $1.14^{+0.06}_{-0.12}$
& $4.07^{+0.18}_{-0.06}$ & $162.5 \pm 1.1 $ \\
\end{tabular}
\end{center}
\end{table}

\begin{table}[]
\caption{Running charged-lepton masses at some typical energy scales
in the SM, including the Higgs mass $M^{}_H \simeq 125~{\rm GeV}$
and the corresponding cutoff scale $\Lambda^{}_{\rm VS} \simeq
4\times 10^{12}~{\rm GeV}$, where the uncertainties of $m^{}_l(\mu)$
are determined by those of $M^{}_l$. Note that the pole masses
$M^{}_l$, rather than the running masses $m^{}_l(M^{}_l)$, are given
in the last row just for comparison.}
\begin{center}
\begin{tabular}{c|c|c|c}
$\mu$ & $m^{}_e(\mu)~ ({\rm MeV})$ &  $m^{}_\mu(\mu)~ ({\rm MeV})$
& $m^{}_\tau(\mu)~ ({\rm MeV})$ \\
\hline
$m^{}_c(m^{}_c)$ & $0.495473903 \pm {0.000000013}$
& $104.4617350^{+0.0000059}_{-0.0000060}$ & $1774.62 \pm 0.16$ \\
\hline
$m^{}_b(m^{}_b)$ & $0.493099926 \pm 0.000000013$
& $103.9961602^{+0.0000059}_{-0.0000060}$ & $1767.02 \pm 0.16$ \\
\hline
$M^{}_W$ & $0.486845781^{+0.000000013}_{-0.000000012}$
& $102.7721083 \pm 0.0000059$ & $1747.05^{+0.15}_{-0.16}$ \\
\hline
$M^{}_Z$ & $0.486570154^{+0.000000012}_{-0.000000013}$
& $102.7181337^{+0.0000059}_{-0.0000058} $ & $1746.17^{+0.15}_{-0.16}$ \\
\hline
$M^{}_H$ & $0.485858771^{+0.000000013}_{-0.000000012}$ &
$102.5788227^{+0.0000058}_{-0.0000059}$ & $1743.89 \pm 0.16$ \\
\hline
$m^{}_t(m^{}_t)$ & $0.485285152^{+0.000000012}_{-0.000000013}$
& $102.4664851^{+0.0000059}_{-0.0000058}$  & $1742.06 \pm 0.16$ \\
\hline
$1~{\rm TeV}$ & $0.489535765^{+0.000000013}_{-0.000000012}$ &
$103.3441945 \pm 0.0000059$ & $1756.81 \pm 0.16$ \\
\hline
$\Lambda^{}_{\rm VS}$ & $0.484511554^{+0.000000012}_{-0.000000013}$
& $102.2835586^{+0.0000058}_{-0.0000059}$ & $1738.82 \pm 0.16$\\
\hline \hline
$M^{}_l$ & $0.510998910 \pm 0.000000013 $  & $105.658367 \pm
0.0000040$
& $1776.82 \pm 0.16$ \\
\end{tabular}
\end{center}
\end{table}


\begin{figure}[]
\vspace{2.0cm}
\epsfig{file=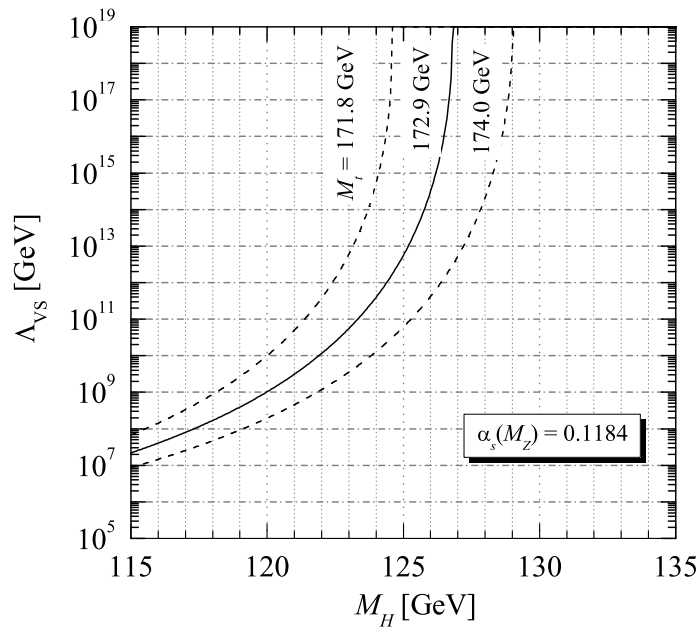,bbllx=-2.5cm,bblly=10.5cm,bburx=2.5cm,bbury=15.5cm,%
width=3.0cm,height=3.0cm,angle=0,clip=0}\vspace{4.0cm}
\caption{Correlation between the energy scale $\Lambda^{}_{\rm VS}$
and the Higgs mass $M^{}_H$ based on the requirement of vacuum
stability, where the solid curve corresponds to the best-fit value
of the top-quark pole mass $M^{}_t = 172.9~{\rm GeV}$, and the
dashed lines stand for the $1\sigma$ lower and upper limits.}
\end{figure}

\begin{figure}[]
\vspace{2.0cm}
\epsfig{file=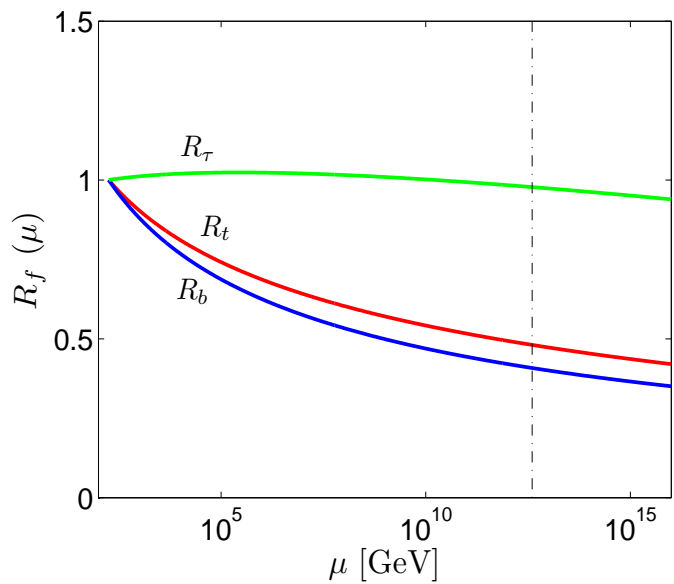,bbllx=-1.1cm,bblly=12cm,bburx=4.1cm,bbury=17cm,%
width=3.5cm,height=3.5cm,angle=0,clip=0}\vspace{3.0cm} \caption{The
running behaviors of $R^{}_t(\mu)$, $R^{}_b(\mu)$ and
$R^{}_\tau(\mu)$ with respect to the energy scale $\mu$ in the SM,
where the vertical dashed line indicates the cutoff scale
$\Lambda^{}_{\rm VS} \simeq 4\times 10^{12}~{\rm GeV}$ as required
by the vacuum stability for $M^{}_H \simeq 125~{\rm GeV}$.}
\end{figure}

\newpage
\begin{figure}[]
\epsfig{file=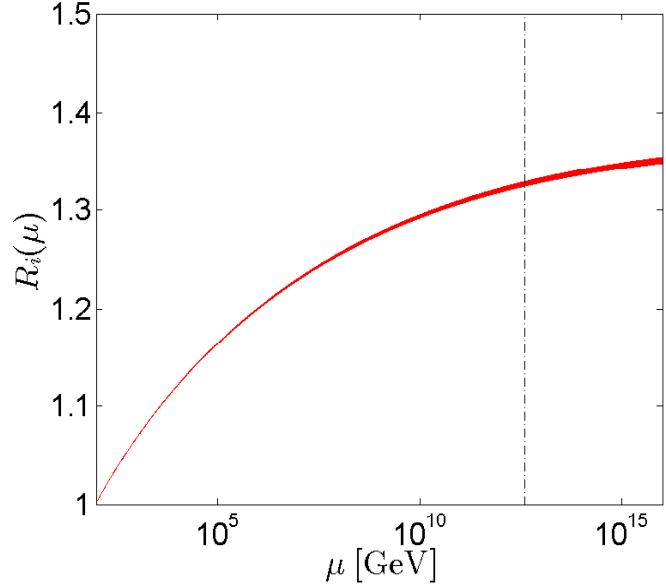,bbllx=-1.1cm,bblly=15cm,bburx=4.1cm,bbury=20cm,%
width=3.5cm,height=3.5cm,angle=0,clip=0}\vspace{5.0cm} \caption{The
evolution of $R^{}_i(\mu)$ with respect to the energy scale $\mu$,
where the red band corresponds to the variation of the Higgs mass in
the range $M^{}_H \simeq (124 \cdots 126)~{\rm GeV}$, and the
vertical dashed line indicates the cutoff scale $\Lambda^{}_{\rm VS}
\simeq 4\times 10^{12}~{\rm GeV}$. Note that $R^{}_1(\mu) \simeq
R^{}_2(\mu) \simeq R^{}_3 (\mu)$ holds to an excellent degree of
accuracy.}
\end{figure}

\begin{figure}[]
\vspace{1.7cm}
\epsfig{file=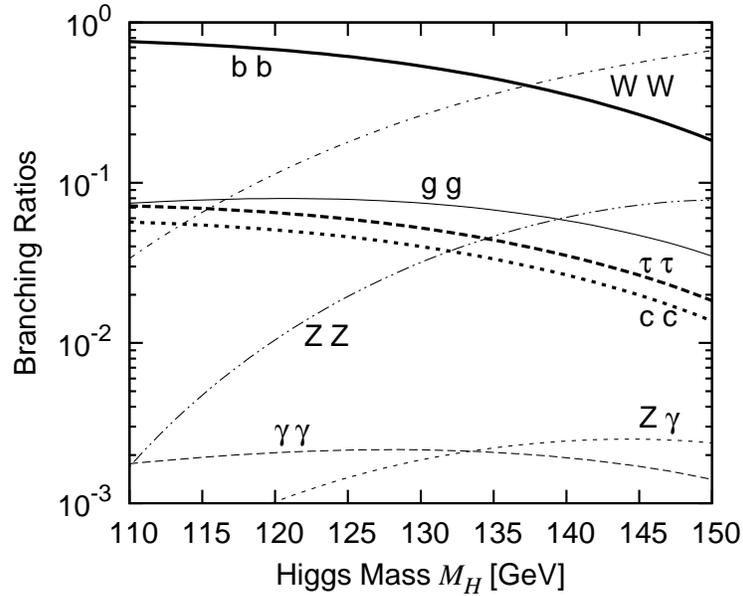,bbllx=0.1cm,bblly=1cm,bburx=5.1cm,bbury=6cm,%
width=8.0cm,height=8.0cm,angle=0,clip=0}\vspace{-1.2cm} \caption{The
branching ratios of two-body Higgs decays versus the Higgs mass
$M^{}_H$. The thick lines stand for the dominant $H \to f \bar{f}$
modes: $b \bar{b}$ (solid line), $\tau^+ \tau^-$ (dashed line) and
$c \bar{c}$ (dotted line); and the thin lines denote $H \to g g$
(solid line), $\gamma \gamma$ (dashed line), $Z \gamma$ (dotted
line), $W^+ W^-$ (dotted-dashed line) and $Z Z$ (double-dotted
dashed line).}
\end{figure}

\end{document}